\documentstyle[12pt]{article}
\begin{document}

\begin{center}{\LARGE\bf
Domain walls in a    Chern-Simons    theory \\ \vskip0.2in
}\end{center}
\begin{center}
by

{   Manuel Torres}\\\vskip0.3in
{\small\it 
 Instituto de F\'isica,   Universidad Nacional  Aut\'onoma de  M\'exico \\  
\vspace{-1mm} 
Apdo. Postal 20-364,  01000  Mexico, D.F., Mexico. \\}

{\bf{Abstract}}
\end{center}
\vskip0.05in
\setlength{\baselineskip}{0.2in}
We study  an   Abelian  Maxwell-Chern-Simons model  in $2 +1 $ dimensions which includes
a magnetic moment interaction. We show that this model possesses domain wall 
as well as vortex solutions. 

\vskip1cm

\section{The model } 

Theories with gauge fields coupled to matter fields in   $(2+1)$ space dimensions present novel 
effects as compared to the   $(3 + 1)$ dimensional case. 
In  planar systems  the Chern-Simons  term (CS) can supplement (sometimes replace) the Maxwell term in the action for the gauge  field \cite{deser} .   Additionally,     the covariant  coupling to the scalar field can be modified by the inclusion of an  extra  term,  which is interpret as an scalar magnetic moment interaction \cite{stern,kogan}.  This  is enforced without  spoiling neither the covariance nor the  gauge invariance of the theory.   The presence of these two terms produce interesting new  effects. We mention   in particular,  that  Chern-Simons solitons carry  electric charge as well as magnetic flux; in addition they possess  fractional spin so they behave as    anyon-like objects  \cite{review}.

We consider  a scalar $QED$ model in $2+1$ dimensions with the  addition of the Chern-Simons term and an anomalous magnetic interaction, it 
 is described by the following effective Lagrangian 
\begin{equation}\label{lag}
 {\cal L} = -{1 \over 4} F_{\mu \nu} F^{\mu \nu} \, + 
   {\kappa \over 4} \epsilon ^{\mu \nu \alpha} A_\mu F_{\nu \alpha}  
\,    + {1 \over 2} |{\cal D}_\mu \phi|^2 -  {1 \over 2} m^2  |\phi | ^2 \, ,
\end{equation}
where  $\kappa$ is the topological mass and   we select for the scalar field
    a simple $\phi^2$ potential. The covariant derivative 
includes both the the  usual minimal coupling plus the magnetic moment interaction 
 \begin{equation}\label{dercov}
{\cal D}_{\mu} = \partial _{\mu} - {ieA_\mu}
- i{g\over 4} \epsilon_{\mu\nu\alpha}F^{\nu\alpha} 
\equiv \partial _{\mu} - {ieA_\mu}
- i{g\over 2} F_{\mu}\, , 
\end{equation}
with $g$ the magnetic moment and we have defined the dual field
$F_\mu =  \epsilon_{\mu \nu \alpha}  F^{\nu \alpha}$.
The possibility of including a magnetic moment 
for scalar particles is a  characteristic property  of the space dimensionality.

There is a limit in which the  gauge field equations reduce
from second- to first-order differential equations 
similar to those of the pure CS theory. Indeed, if 
the relation   $\kappa = - {2 e \over g}$ holds the equation of motion for 
the gauge field is given by 
\begin{equation}\label{pcs}
\kappa  F_\mu  =    J_\mu
\, ,  \end{equation}
where $J_\mu$ is the conserved Noether current. 
This equation of motion  implies  that any object carrying magnetic flux ($  \Phi_B$) must
also carry electric charge ($Q$), with the two quantities related
as  $Q =  \kappa   \Phi_B $. 
If we work within the pure CS limit a  self-dual  Maxwel-Chern-Simons  gauge model 
can be constructed  \cite{torres1}.   This is attained if  the scalar potential has the particular 
 $\phi^2$ form   and the scalar mass is made equal to the  CS mass; 
the energy obeys a Bogomol'nyi  lower bound that is saturated by the fields that satisfy the 
self-duality equations. The vortex solutions of this self-dual model have been studied in detail 
\cite{torres2,torres3}.
Additionally,  it is also possible to find  domain wall configurations for this model.

\section{Domain walls} 

The  present $\phi^2$  theory possesses  a single minimum, yet  it is possible  to find  one dimensional  soliton solutions of the    domain wall type. Consider a one dimensional structure  depending only on   the $x$  variable,  both  at $x \to \infty$  and $x \to - \infty$ the scalar field  should vanish. However,  there can be  an intermediate region where  $\phi \neq 0$, $i.e.$, a region of false vacuum.
The  maximum of $\phi$ determines the position of the  wall.
  The domain wall  carries both magnetic flux and electric charge 
per unit length.  Seeking  a domain   wall solution parallel to the $y$-axis, the translational 
invariance of the theory implies that all the fields depend only on $x$.
By an appropriate gauge transformation the scalar field is made real everywhere
$ \phi = (\kappa/e) f$ and the  gauge potential $\vec A$ is selected along the $y$ axis. 
Using the pure CS equations of motion  (\ref{pcs}), 
 the  expression   for the energy   can be written as 
\begin{eqnarray}\label{ew1}
E = { 1 \over 2}  \int  d^2 r  \bigg[  
 \left({\left(1 - f^2  \right)^{1/2} \over f } {d A_y\over d x} 
\,\, \mp \,\, {\kappa f A_y \over \left(1 - f^2  \right)^{1/2}  }  \right)^2
\nonumber \\
 \,\,+ \,\,  {\kappa^2 \over e^2} \left({d f \over d x} \, \, \pm \, \, m f   \right)^2
  \,\,\mp \,\, {m \kappa^2 \over e^2}{d f^2\over d x}
\pm \kappa {d  A_y^2\over d x}
\bigg]
\, . 
\end{eqnarray}
The boundary conditions for the scalar field
are $f(-\infty) = f(\infty) = 0$. The magnetic flux  per unit length $(\gamma)$
is given by  $\gamma = A_y(\infty)  - A_y(-\infty)$, so 
$A_y(\infty)  \neq A_y(-\infty)$  is required in order to get a non-vanishing magnetic flux.  A configuration is sought which   has a definite symmetry  with respect to the 
position $X$ of the domain wall, then    
$A_y(\infty)  = -  A_y(-\infty) \equiv \gamma/2$ is selected. 

The static solution is obtained  minimizing the energy per unit length  with 
$\gamma$ fixed. The boundary conditions cannot be satisfied if   the same upper (or  lower) signs  in  (\ref{ew1}) are used  for all    $x$.  Rather,  the upper signs in the region to the  right  of the domain wall ($x > X$) are selected, whereas  for
 $x < X$ we take the lower signs.
With this selection the minimum energy per unit length becomes 
\begin{equation}\label{ew2}
{\cal E} = {\kappa^2 m \over e^2 } f_0^2  +  { \kappa \over 4} \gamma^2 
\, , 
\end{equation}
where $f_0 \equiv f(X)$. This result is obtained provided that the fields
satisfy the following  equations:
\begin{eqnarray}\label{edw}
{df \over d x} &=& \mp m f \,  , \nonumber \\
{d A_y \over d x} = &=& \pm {\kappa f ^2 \over \left( 1 - f^2  \right) } A_y
\, , 
\end{eqnarray}
where the upper (lower) sign must be taken for $ x > X$  ($ x <  X$).
These equations  are easily integrated to give 
\begin{eqnarray}\label{sw}
f(x ) &=& e^{- m |x - X|}  \, ,  \nonumber \\
A_y(x) &=&  sgn(x - X) {\gamma \over 2}\left(1 - e^{-2 m |x - X|}  \right)^{\kappa/2 m }    \, . 
\end{eqnarray}
This is   a domain wall configuration   localized at   $x = X$  with a width 
of order $1/m$.  The solution to the  first equation in (\ref{edw}) does  not restrict the value   of $f_0$. However,  $f_0 = 1$ has to  set so 
the gauge field be continuous everywhere. The anti-kink configuration is obtained  by simply reversing   the signs of the fields in (\ref{sw}).

The domain wall carries a magnetic flux and charge per unit length
given by $\gamma $ and  $- \kappa \gamma$ respectively.
The  magnetic field is given by 
\begin{equation}\label{mfw}
B= {\kappa  \gamma \over 2}  e^{- m |x - X|} 
 \left(1 - e^{-2 m |x - X|}  \right)^{\kappa/2 m - 1} \, .
\end{equation}
Notice  that for $\kappa < 2 m$ the magnetic field is concentrated near 
$ x = X$ and falls off rapidly away from the wall. Instead for 
$\kappa  \geq 2 m$ the  magnetic field vanishes at  $x = X$ and
the profile  of $B$ is  doubled peaked  with maximums  
 at $x = X \pm {1 \over  m} | \ln \left({\kappa \over 2 m} \right)|$. 

 To investigate the conditions required to have stable configurations consider the decay of the
domain wall  by  the emission of  scalar particles.  Because of charge conservation 
a decaying wall should radiate $\kappa \gamma / e$  quantas of scalar particles per unit length. Thus,  the energy of the elementary excitations per unit length  at rest will be
 $\kappa \gamma  m / e$.  The stability condition  requires   this energy
 to be  bigger  than the soliton energy in
 (\ref{ew2}):
\begin{equation}\label{ew3}
 {\kappa^2 m \over e^2 }   +  { \kappa \over 4} \gamma^2 
\leq   { \kappa \gamma  m \over  e }
\, . 
\end{equation}
This condition  implies $ \kappa < m$ and also yields both an  upper and a lower 
bound to the  values of the  magnetic flux 
per unit length: $ \gamma_-  \leq \gamma \leq \gamma_+$, where
\begin{equation}\label{eque}
\gamma_\pm =  {2 m \over e } \left[ 1 \pm \sqrt{1 - { \kappa \over m   }}\right ]
\, , 
\end{equation}

The domain wall solutions  in  (\ref{sw})  can  be utilized   to  study rotationally symmetric  configurations of the vortex type. 
The vortex configurations simplify in the large vorticity  limit 
In this large$-n$ limit the vortex can be considered as   a     ring  of  large radius
 $R \approx n/\kappa$   and  thickness
of order $1 /m$  separating  two regions  of vacuum. 
The magnetic flux is concentrated
within this domain of  width $\sim  1/ m$   where   a region of false vacuum  ($\phi \neq 0$) is trapped. In this limit   $R \gg1 /m \sim 1/\kappa$, and  the fields near the
ring is  well approximated by the domain wall solution  \cite{torres3}.

This model raises a number of  interesting questions for further investigation.   
In particular  a complete description of the multisoliton solution  deserve to be clarified. 
It may also be interesting to investigate   the properties of the Chern-Simons vortices  and domain walls upon quantization. 

\vskip1.cm
 I acknowledge support from  UNAM (proyecto DGAPA IN103895) \\
and  CONACyT (proyecto  3097 p-E).

{99}

\end{document}